# Complete symmetry analyses of the surface-induced piezomagnetic, piezoelectric and linear magnetoelectric effects

E.A. Eliseev

Institute for Problems of Materials Science, NAS of Ukraine,

Krjijanovskogo 3, 03142 Kiev, Ukraine,

The impact of the surface on the magnetic symmetry is studied. It is shown how symmetry breaking transforms the "seeding" 90 bulk magnetic classes, which include 66 piezomagnetic classes and 58 magnetoelectric classes, into the 21 surface-induced magnetic classes, which all are piezomagnetic, piezoelectric and 19 of them are linear magnetoelectric. So, the new piezomagnetic, piezoelectric properties and linear magnetoelectric coupling should appear in nanomaterials, which are absent in the bulk.

**PACS:** 77.80.-e, 77.84.Dy, 68.03.Cd, 68.35.Gy

**Keywords:** symmetry theory, piezomagnetic, piezoelectric, linear magnetoelectric effect

## I. Introduction

The properties of nanomaterials absent in the bulk (antiferroelecric-ferroelectric and antiferromagnetic-ferromagnetic size-induced transitions, self-polarization and self-magnetization, superparamagnetism, superparaelctricity) are the most interesting from both the fundamental and applied physics point of views. Since a translational symmetry could be broken at any surface or interface, intriguing modifications of the structure polar, magnetic and electronic state typically appeared in thin films and nanoparticles [1, 2, 3, 4, 5, 6, 7, 8].

It was shown earlier (see Refs. [9, 10, 11]), that the possible physical origin of electrical polarization (or magnetization) in nonferroelectric (or nonferromagnetic) nanomaterials could be mechanical conditions at their surface. Surface-induced (i.e. built-in) electric field leads to the horizontal shift of all hysteresis loops and electret-like state appearance in ferroelectric films with thickness less than the critical one. It facilitates thin film the self-polarization as predicted in Refs.[12, 13]. Beside the hysteresis loop horizontal shifts, the ordering effects caused by the surface-induced magnetic fields should induce ferromagnetism or irreversible magnetization in nanosystems absent in the bulk material [11].

The magnetoelectric (**ME**) effect, i.e. when the application of either a magnetic field or an electric field induces an electric polarization as well as magnetization, attracted much attention in the last years [14]. Recently the revival of the ME effect has been observed due to discovery of high (several hundred percents) ME effects both in single phase and composite materials (see [15], [16, 17], [18] review [19] and ref. therein). Most of the composites exhibit high extrinsic ME effect resulting from the interaction between magnetic and electric components via e.g. their magnetostriction and piezoelectric properties, as well as piezomagnetic-piezoelectric interaction [20, 21, 22, 23]. However no indication on the possibility to obtain high ME coupling was revealed for conventional type of mechanical conditions in the bulk materials. The situation becomes completely different under the confinement of nanomaterials. Some evidence in favour of this assumption follows from the observation of dramatically higher than in the bulk crystals ME coefficients in epitaxially (001) oriented $BiFeO_3$ films on a $SrTiO_3$ substrate. This effect was related with the influence of boundary conditions in the consideration performed in [24].

Using symmetry theory Eliseev et al [25] explore how symmetry breaking inevitably present in the vicinity of any surface gives rise to spontaneous surface piezomagnetic, piezoelectric and magnetoelectric effects in the nanomaterials of the bulk m3m, m′3m′, m′3m and m3m′ cubic magnetic symmetry groups The surface 4mm, 4m′m′, 4′m′m and 4′mm′ symmetry groups were directly obtained from the bulk m3m, m′3m′, m′3m and m3m′ symmetry groups respectively by considering the surface normal $x_3$ ↑↑ 4 (symbol 4 stands for the forth order rotation axis). In bulk materials the piezomagnetic effect could exist in 66 magnetic classes [26]. In the paper we perform the complete symmetry analyses for all existing 90 bulk magnetic classes.

**II. Transformation laws**

By definitions, when one changes system of coordinates the components of tensors change accordingly. The transformation laws for the axial ($g^{(m)}_{lpn}$), polar ($g^{(e)}_{lmn}$) third rank tensors describing piezomagnetic (*m*) and piezoelectric (*e*) effects, and axial tensor of linear magnetoelectric effect in bulk are [26]:

$$\widetilde{g}^{(m)}_{ijk} = (-1)^{tr} \det(\mathbf{A}) A_{il} A_{jp} A_{kn} g^{(m)}_{lpn}, \tag{1a}$$

$$\widetilde{g}^{(e)}_{ijk} = A_{il} A_{jm} A_{kn} g^{(e)}_{lmn}. \tag{1b}$$

$$\widetilde{\gamma}_{ij} = (-1)^{tr} \det(\mathbf{A}) A_{ik} A_{jl} \gamma_{kl} \tag{1c}$$

Here **A** is the transformation matrix with components $A_{ij}$ ($i,j = 1,2,3$), relating two coordinate systems with coordinates $\tilde{x}_i$ and $x_j$, namely $\tilde{x}_i = A_{ij} x_j$. We restrict the charges of coordinate to the rotations, reflections and inversion, in this case $\det(\mathbf{A}) = \pm 1$. The factor *tr* denotes either the presence (*tr* = 1) or the absence (*tr* = 0) of time-reversal operation coupled to the entirely space point transformation $A_{ij}$. For case when **A** is the transformation matrix representing the element of the material point symmetry group (considered hereinafter) the identity $\tilde{g}_{ijk}^{(m,e)} \equiv g_{ijk}^{(m,e)}$ should be valid.

For any spatially confined system the inversion center disappears in surface normal direction and only the symmetry axes and planes normal to the surface conserves near the surface [27]. Thus the magnetic and space symmetry group should be reduced to one of its subgroup, consisting of the transformation matrices $A_{ij}^S$, which satisfy the relations $n_i A_{ij}^S n_j = 1$ where $n_j$ are the components of the surface normal. For instance, for the surface normal parallel to $x_3$, the rotations and mirror reflections matrices must have the form $\mathbf{A}^S = \begin{pmatrix} a_{11} & a_{12} & 0 \\ a_{21} & a_{22} & 0 \\ 0 & 0 & 1 \end{pmatrix}$, at that $\det(\mathbf{A}^S) = \det(\mathbf{A}) = \pm 1$. As a result the surface piezoeffect tensors $g_{lpn}^{(Sm)}$ and $g_{lmn}^{(Se)}$ the surface linear magnetoelectric effect tensor $\gamma_{ij}^S$ should obey another transformation laws than the ones existing in the bulk of material, namely $g_{ijk}^{(Sm)} \equiv (-1)^{tr} \det(\mathbf{A}^S) A_{il}^S A_{jp}^S A_{kn}^S g_{lpn}^{(Sm)}$, $g_{ijk}^{(Se)} \equiv A_{il}^S A_{jm}^S A_{kn}^S g_{lmn}^{(Se)}$ and $\gamma_{ij}^S = (-1)^{tr} \det(\mathbf{A}) A_{ik} A_{il} \gamma_{kl}^S$.

**III. Linear magnetoelectric and piezomagnetic properties of the surface-induced magnetic classes**

The new piezomagnetics, piezoelectrics and linear magnetoelectrics should appear even for nanosystems from materials which are nonpiezomagnetic and nonpiezoelectric in the bulk form [25]. To explore the possibility, we calculate the evident form of the surface magnetic symmetry groups for all 90 bulk magnetic symmetry groups belonging to the six syngonies. Results of the symmetry consideration are presented in the **Tables 1-6**, where the first left column is the bulk magnetic class (one of the 90 existing), next three (or five) columns are the bulk material piezoelectric, linear ME, piezomagnetic, ferroelectric and ferromagnetic (if any) properties. The last three right columns are the **surface induced classes** calculated from the corresponding **bulk** **class** by the direct inspection method. Note that the prime "′" near the symmetry element means its combination with **time inversion**.

### Tables 1. Triclinic syngony

| Bulk class | ferro-magnetic | linear ME | Piezo-magnetic | ferro-electric | piezo-electric | Surface induced classes for any surface |
|---|---|---|---|---|---|---|
| 1 | yes | yes | yes | yes | yes | 1 |
| $\bar{1}$ | yes | no | yes | no | no | 1 |
| $\bar{1}'$ | no | yes | no | | | 1 |

### Tables 2. Monoclinic syngony

| Bulk class | ferro-magnetic | linear ME | piezo-magnetic | ferro-electric | piezo-electric | Surface induced classes | | |
|---|---|---|---|---|---|---|---|---|
| | | | | | | (100) | (010) | (001) |
| 2 | yes | yes | yes | yes | yes | 1 | 1 | 2 |
| 2′ | yes | yes | yes | | | 1 | 1 | 2′ |
| *m* | yes | yes | yes | yes | yes | *m* | *m* | 1 |
| *m*′ | yes | yes | yes | | | *m*′ | *m*′ | 1 |
| 2/*m* | yes | no | yes | no | no | *m* | *m* | 2 |
| 2/*m*′ | no | yes | no | | | *m*′ | *m*′ | 2 |
| 2′/*m* | no | yes | no | | | *m* | *m* | 2′ |
| 2′/*m*′ | yes | no | yes | | | *m*′ | *m*′ | 2′ |

### Tables 3. Rhomb syngony

| Bulk magnetic class | ferro-magnetic | linear ME | piezo-magnetic | ferro-electric | piezo-electric | Surface induced magnetic classes | | |
|---|---|---|---|---|---|---|---|---|
| | | | | | | (100) | (010) | (001) |
| 222 | no | yes | yes | no | yes | 2 | 2 | 2 |
| 22′2′ | yes | yes | yes | | | 2′ | 2′ | 2 |
| *mm*2 | no | yes | yes | yes | yes | 1 | 1 | *mm*2 |
| *m*′*m*′2 | yes | yes | yes | | | 1 | 1 | *m*′*m*′2 |
| *mm*′2′ | yes | yes | yes | | | 1 | 1 | *mm*′2′ |
| *mmm* | no | no | yes | no | no | *mm*2 | *mm*2 | *mm*2 |
| *m*′*m*′*m*′ | no | yes | no | | | *m*′*m*′2 | *m*′*m*′2 | *m*′*m*′2 |
| *mmm*′ | no | yes | no | | | *mm*′2′ | *mm*′2′ | *mm*2 |
| *m*′*m*′*m* | yes | no | yes | | | *mm*′2′ | *mm*′2′ | *m*′*m*′2 |

**Tables 4. Tetragonal syngony**

| Bulk class | ferro-magnetic | linear ME | piezo-magnetic | Surface induced magnetic classes | | |
|---|---|---|---|---|---|---|
| | | | | (100) | (010) | (001) |
| 4 | yes | yes | yes | 1 | 1 | 4 |
| 4' | no | yes | yes | 1 | 1 | 4' |
| $\bar{4}$ | yes | yes | yes | 1 | 1 | 2 |
| $\bar{4}'$ | no | yes | yes | 1 | 1 | 2 |
| $\bar{4}2m$ | no | yes | yes | m | m | mm2 |
| $\bar{4}2'm'$ | yes | yes | yes | m' | m' | m'm'2 |
| $\bar{4}'2m'$ | no | yes | yes | m' | m' | m'm'2 |
| $\bar{4}'2'm$ | no | yes | yes | m | m | mm2 |
| 422 | no | yes | yes | 2 | 2 | 4 |
| 42'2' | yes | yes | yes | 2' | 2' | 4 |
| 4'22' | no | yes | yes | 2 | 2 | 4' |
| 4/m | yes | no | yes | 1 | 1 | 4 |
| 4/m' | no | yes | no | 1 | 1 | 4 |
| 4'/m | no | no | yes | 1 | 1 | 4' |
| 4'/m' | no | yes | no | 1 | 1 | 4' |
| 4mm | no | yes | yes | m | m | 4mm |
| 4m'm' | yes | yes | yes | m' | m' | 4m'm' |
| 4'mm' | no | yes | yes | m | m | 4'mm' |
| 4/m mm | no | no | yes | mm2 | mm2 | 4mm |
| 4/m' m'm' | no | yes | no | m'm'2 | m'm'2 | 4m'm' |
| 4/m m m' | yes | no | yes | mm2' | mm'2' | 4m'm' |
| 4/m' mm | no | yes | no | mm'2' | mm'2' | 4mm |
| 4'/m mm' | no | no | yes | mm2 | mm2 | 4'mm' |
| 4'/m' mm' | no | yes | no | mm'2' | mm'2' | 4'mm' |

**Tables 5. Hexagonal syngony**

| Bulk magnetic class | Ferro-magnetic | Linear ME | Piezo-magnetic | Surface induced magnetic classes | | |
|---|---|---|---|---|---|---|
| | | | | $(11\bar{2}0)$ | $(0\bar{1}10)$ | $(0001)$ |
| 3 | yes | yes | yes | 1 | 1 | 3 |
| 32 | no | yes | yes | 1 | 2 | 3 |
| 32' | yes | yes | yes | 1 | 2' | 3 |
| 3m | no | yes | yes | m | 1 | 3m |
| 3m' | yes | yes | yes | m' | 1 | 3m' |
| $\bar{3}$ | yes | no | yes | 1 | 1 | 3 |
| $\bar{3}'$ | no | yes | no | 1 | 1 | 3 |
| $\bar{3}m$ | no | no | yes | m | 2 | 3m |
| $\bar{3}m'$ | yes | no | yes | m' | 2' | 3m' |
| $\bar{3}'m'$ | no | yes | no | m' | 2 | 3m' |
| $\bar{3}'m$ | no | yes | no | m | 2' | 3m |
| $\bar{6}$ | yes | no | yes | 1 | 1 | 3 |
| $\bar{6}'$ | no | yes | yes | 1 | 1 | 3 |
| $\bar{6}m2$ | no | no | yes | 1 | mm2 | 3m |
| $\bar{6}m'2'$ | yes | no | yes | 1 | mm'2' | 3m' |
| $\bar{6}'m'2$ | no | yes | yes | 1 | m'm'2 | 3m' |
| $\bar{6}'m2'$ | no | yes | yes | 1 | mm'2' | 3m |
| 6 | yes | yes | yes | 1 | 1 | 6 |
| 6' | no | no | yes | 1 | 1 | 6' |
| 622 | no | yes | yes | 2 | 2 | 6 |
| 62'2' | yes | yes | yes | 2' | 2' | 6 |
| 6'22' | no | no | yes | 2 | 2' | 6' |
| 6/m | yes | no | yes | m | m | 6 |
| 6/m' | no | yes | no | m' | m' | 6 |
| 6'/m' | no | no | yes | m' | m' | 6' |
| 6'/m | no | no | no | m | m | 6' |
| 6mm | no | yes | yes | m | m | 6mm |
| 6m'm' | yes | yes | yes | m' | m' | 6m'm' |
| 6'mm' | no | no | yes | m | m' | 6'mm' |
| 6/mmm | no | no | yes | mm2 | mm2 | 6mm |
| 6/m'mm | no | yes | no | mm'2' | mm'2' | 6mm |
| 6/m'm'm' | no | yes | no | m'm'2 | m'm'2 | 6m'm' |
| 6/mm'm' | yes | no | yes | mm'2' | mm'2' | 6m'm' |
| 6'/mmm' | no | no | no | mm2 | mm'2' | 6'mm' |
| 6'/m'mm' | no | no | yes | mm'2' | m'm'2 | 6'mm' |

**Tables 6. Cubic syngony**

| Bulk magnetic class | Ferro-magnetic | Linear ME | Piezo-magnetic | Surface induced magnetic classes |||
|---|---|---|---|---|---|---|
| | | | | (100) | (110) | (111) |
| 23 | no | yes | yes | 2 | 1 | 3 |
| $m3$ | no | no | yes | $mm2$ | $m$ | 3 |
| $m'3$ | no | yes | no | $m'm'2$ | $m'$ | 3 |
| $\bar{4}3m$ | no | no | no | $mm2$ | $m$ | $3m$ |
| $\bar{4}'3m'$ | no | yes | yes | $m'm'2$ | $m'$ | $3m'$ |
| 432 | no | yes | no | 4 | 2 | 3 |
| $4'32'$ | no | no | yes | $4'$ | $2'$ | 3 |
| $m3m$ | no | no | no | $4mm$ | $mm2$ | $3m$ |
| $m'3m'$ | no | yes | no | $4m'm'$ | $m'm'2$ | $3m'$ |
| $m'3m$ | no | no | no | $4'mm'$ | $mm'2'$ | $3m$ |
| $m3m'$ | no | no | yes | $4'mm'$ | $mm'2'$ | $3m'$ |

Using the **Tables 1-6**, the evident form of surface and bulk piezoelectric, piezomagnetic and ME tensors can be extracted from the well-known tables of e.g. Ref. [26] for the definite symmetry group listed in the right columns entitled "surface-induced magnetic classes".

Sometimes more valuable are the "inverse" information: Such information is summarized in the **Table 7**, where the first left column is the surface-induced magnetic class (one of the 90 existing), next three columns are the surface-induced ferromagnetic, linear ME and piezomagnetic properties. The "seeding" bulk magnetic classes, which definite surface cuts (right column) give the surface class in the first left column, are listed in the middle column.

Direct analyses of the **Table 7** show that all 21 surface-induced magnetic classes are piezomagnetic and 19 of them are magnetoelectric, while the seeding 90 bulk magnetic classes include 66 piezomagnetic classes and 58 magnetoelectric classes.

Piezoelectric, piezomagnetic and ME tensors, which correspond to the 21 surface-induced magnetic classes, can be extracted from the well-known tables e.g. of Ref. [26]. Thus it follows from the **Table 7** that nonpiezoelectric and nonpiezomagnetic bulk material with magnetic symmetry becomes piezoelectric and piezomagnetic in the vicinity of surface with different $g_{ijk}^{(Sm)}$ tensors, which depend on the surface symmetry group. Piezomagnetic bulk materials remain piezomagnetic in the vicinity of surface, but the symmetry of $g_{ijk}^{(Sm)}$ changes in comparison with a bulk tensor $g_{ijk}^{(m)}$. Except bulk classes $6'mm'$, $6'/mmm'$, $6'/m'mm'$, $6'$, $6'22'$, $6'/m'$, $6'/m$, all other non-magnetoelectric bulk materials become linear ME in the vicinity of surface with different $\gamma_{ij}^S$ tensors, which depend on the surface symmetry group. The linear magnetoelectric coupling should dramatically changes the phase diagrams of ferroic nanosystems.

**Table 7.** Ferromagnetic (FM), linear magnetoelectric (ME) and piezomagnetic (PM) properties of the surface magnetic class (left columns), which are obtained from the bulk magnetic classes (middle column) by definite surface cuts (right column)

| Surface magnetic class | FM | ME | PM | PE | Bulk magnetic class | surfaces (cuts) |
|---|---|---|---|---|---|---|
| $6mm$ | no | yes | yes | yes | $6mm$, $6/mmm$, $6/m'mm$ | (001) |
| $6m'm'$ | yes | yes | yes | yes | $6m'm'$, $6/m'm'm'$, $6/mm'm'$ | |
| $6'mm'$ | no | no | yes | yes | $6'mm'$, $6'/mmm'$, $6'/m'mm'$ | |
| $6$ | yes | yes | yes | yes | $6, 622, 62'2', 6/m, 6/m'$, | (001) |
| $6'$ | no | no | yes | yes | $6', 6'22', 6'/m', 6'/m$ | |
| $3m$ | no | yes | yes | yes | $3m$, $\bar{3}m$, $\bar{3}'m$, $\bar{6}m2$, $\bar{6}'m2'$, | (001) |
| | | | | | $\bar{4}3m$, $m3m$, $m'3m$ | (111) |
| $3m'$ | yes | yes | yes | yes | $3m'$, $\bar{3}m'$, $\bar{3}'m'$, $\bar{6}m'2'$, | (001) |
| | | | | | $\bar{6}'m'2$, $\bar{4}'3m'$, $m'3m'$, | |
| | | | | | $m3m'$ | (111) |
| $3$ | yes | yes | yes | yes | $3, 32, 32', \bar{3}, \bar{3}', \bar{6}, \bar{6}'$ | (001) |
| | | | | | $23, m3, m'3, 432, 4'32'$ | (111) |
| $4mm$ | no | yes | yes | yes | $4mm$, $4/mmm$, $4/m'mm$ | (001) |
| | | | | | $m3m$ | (100), (010), (001) |
| $4m'm'$ | yes | yes | yes | yes | $4m'm'$, $4/m'm'm'$, $4/mm'm'$ | (001) |
| | | | | | $m'3m'$ | (100), (010), (001) |
| $4'mm'$ | no | yes | yes | yes | $4'mm'$, $4'/mmm'$, $4'/m'mm'$ | (001) |
| | | | | | $m'3m$, $m3m'$ | (100), (010), (001) |
| $4$ | yes | yes | yes | yes | $4, 422, 42'2', 4/m, 4/m'$ | (001) |
| | | | | | $432$ | (100), (010), (001) |
| $4'$ | no | yes | yes | yes | $4', 4'22', 4'/m, 4'/m'$ | (001) |
| | | | | | $4'32'$ | (100), (010), (001) |
| $mm2$ | no | yes | yes | yes | $mm2, mmm', \bar{4}'2'm, \bar{4}2m$, | (001) |
| | | | | | $4/mmm$, $4'/mmm'$ | (010), (100) |
| | | | | | $mmm$, $m3$, $\bar{4}3m$ | (100), (010), (001) |
| | | | | | $m3m$ | (110), (101), (011) |
| | | | | | $\bar{6}m2$, $6/mmm$ | $(0\bar{1}10)$ |
| | | | | | $6'/mmm'$, $6/mmm$ | $(11\bar{2}0)$ |
| $m'm'2$ | yes | yes | yes | yes | $m'm'2, m'm'm, \bar{4}2'm', \bar{4}'2m'$ | (001) |
| | | | | | $4/m'm'm'$ | (010), (100) |
| | | | | | $m'm'm'$, $m'3$, $\bar{4}'3m'$ | (100), (010), (001) |
| | | | | | $m'3m'$ | (110), (101), (011) |
| | | | | | $\bar{6}'m'2$, $6/m'm'm'$, $6'/m'mm'$ | $(0\bar{1}10)$ |
| | | | | | $6/m'm'm'$ | $(11\bar{2}0)$ |
| $mm'2'$ | yes | yes | yes | yes | $mm'2'$ | (001) |
| | | | | | $mmm'$, $m'm'm$, $4/mm'm'$, $4/m'mm$, $4'/m'mm'$ | (010), (100) |
| | | | | | $m'3m$, $m3m'$ | (110), (101), (011) |
| | | | | | $\bar{6}m'2'$, $\bar{6}'m2'$, $6/m'mm$, $6/mm'm'$, $6'/mmm'$ | $(0\bar{1}10)$ |
| | | | | | $6/m'mm$, $6/mm'm'$, $6'/m'mm'$ | $(11\bar{2}0)$ |
| 2 | yes | yes | yes | yes | $2$, $2/m$, $2/m'$, $22'2'$, $\bar{4}$, $\bar{4}'$ | (001) |
| | | | | | $422$, $4'22'$ | (100), (010) |
| | | | | | $222$, $23$ | (100), (010), (001) |
| | | | | | $432$ | (110), (101), (011) |
| | | | | | $32$, $\bar{3}m$, $\bar{3}'m'$, $622$ | $(0\bar{1}10)$ |
| | | | | | $622$, $6'22'$ | $(11\bar{2}0)$ |
| $2'$ | yes | yes | yes | yes | $2'$, $2'/m$, $2'/m'$ | (001) |
| | | | | | $22'2'$, $42'2'$ | (100), (010) |
| | | | | | $4'32'$ | (110), (101), (011) |
| | | | | | $32'$, $\bar{3}m'$, $\bar{3}'m$, $62'2'$, $6'22'$ | $(0\bar{1}10)$ |
| | | | | | $62'2'$ | $(11\bar{2}0)$ |
| $m$ | yes | yes | yes | yes | $m$, $2/m$, $2'/m$, $\bar{4}2m$, $\bar{4}'2'm$, $4mm$, $4'mm'$, | (100), (010) |
| | | | | | $6/m$, $6'/m$, $6mm$, | $(0\bar{1}10)$ |
| | | | | | $3m$, $\bar{3}m$, $\bar{3}'m$, $6/m$, $6'/m$, $6mm$, $6'mm'$ | $(11\bar{2}0)$ |
| | | | | | $m3$, $\bar{4}3m$ | (110), (101), (011) |
| $m'$ | yes | yes | yes | yes | $m'$, $2/m'$, $2'/m'$, $\bar{4}2'm'$, $\bar{4}'2m'$, $4m'm'$, | (100), (010) |
| | | | | | $6/m'$, $6'/m'$, $6m'm'$, $6'mm'$ | $(0\bar{1}10)$ |
| | | | | | $3m'$, $\bar{3}m'$, $\bar{3}'m'$, $6/m'$, $6'/m'$, $6m'm'$ | $(11\bar{2}0)$ |
| | | | | | $m'3$, $\bar{4}'3m'$ | (110), (101), (011) |
| 1 | yes | yes | yes | yes | $1$, $\bar{1}$, $\bar{1}'$ | Any surface* |

*Note, that for other bulk classes (with higher symmetry) surfaces different from listed above also could lead to the complete disappearance of symmetry (reduction to class 1).

**Summary**

We calculated how symmetry breaking inevitably present in the vicinity of any surface transforms the "seeding" 90 bulk magnetic classes, which include 66 piezomagnetic classes and 58 magnetoelectric classes, into the 21 surface-induced magnetic classes, which all are piezomagnetic, piezoelectric and 19 of them are linear magnetoelectric. Thus piezomagnetic, piezoelectric properties and linear magnetoelectric coupling should appear in nanomaterials (e.g. ultrathin films, nanocomposites and nanopowders), which are absent in the bulk.

Since our predictions are based on the symmetry considerations, we should underline that although the surface-induced symmetry lowering proclaims the possibility of the magnetic moments appearance at the surface, a separate first principles based study is required to establish the microscopic physical mechanisms, which lead to the magnetism origin in every concrete case.

**Acknowledgement**

Author is very grateful to Maya D. Glinchuk and Anna N. Morozovska for useful remarks and multiple discussions